\documentclass[a4paper]{article}

\usepackage{INTERSPEECH2021}
\usepackage{multirow}
\usepackage{cite}
\usepackage{hyperref}
\hypersetup{hidelinks}
\title{PercepNet+: A Phase and SNR Aware PercepNet for Real-Time Speech Enhancement}
\name{Xiaofeng Ge$^1$, Jiangyu Han$^1$, Yanhua Long$^{1,2*}$\thanks{Yanhua Long is the corresponding author. The work is supported by the National Natural Science Foundation of China (Grant No.62071302).}, Haixin Guan$^2$}
\address{
  $^1$Shanghai Normal University, Shanghai, China\\
  $^2$Unisound AI Technology Co., Ltd., Beijing, China}
\email{\{xfge01,jyhan03\}@163.com, yanhua@shnu.edu.cn, guanhaixin@unisound.com}

\begin{document}

\maketitle
\begin{abstract}

PercepNet, a recent extension of the RNNoise, an efficient,
high-quality and real-time full-band
speech enhancement technique, has shown promising
performance in various public deep noise suppression tasks. This paper
proposes a new approach, named PercepNet+, to further extend the
PercepNet with four significant improvements.
First, we introduce a phase-aware structure to leverage the phase
information into PercepNet,
by adding the complex  features
and complex subband gains as the deep network input and output respectively.
Then, a signal-to-noise ratio (SNR) estimator and an SNR-switched post-processing are specially designed to
alleviate the over attenuation (OA) that appears in high SNR conditions
of the original PercepNet. Moreover, the GRU layer is replaced by
TF-GRU to model both temporal and frequency dependencies. Finally,
we propose to integrate the loss of complex subband gain, SNR, pitch filtering
strength, and an OA loss in a multi-objective learning manner to further
improve the speech enhancement performance.
Experimental results show that, the proposed PercepNet+ outperforms
the original PercepNet significantly in terms of both PESQ and STOI,
without increasing the model size too much.

\end{abstract}


\vspace{-0.2cm}
\section{Introduction}
\label{sec:intro}

Speech enhancement (SE) aims to improve speech perceptually quality and intelligibility under noisy condition..
Recently, deep learning based SE approaches \cite{XuDDL15,WangNW14}
have shown performance over most traditional methods, such as log-spectral amplitude estimation \cite{lsa85},
spectral subtraction \cite{1979Suppression}, etc. In many scenarios,
such as telecommunication and online conference, SE systems are required
to meet both good denoising performance and real-time constrains.
For real-time SE, the current mainstream methods can be divided into two categories.
One is end-to-end systems based on U-Net structure \cite{RonnebergerFB15,DENSE18},
such as DCCRN \cite{HuLLXZFWZX20}, DCCRN+ \cite{lv21_interspeech}, and DPCRN \cite{le21b_interspeech}, etc.
The other is perceptually-motivated, hybrid signal processing/deep learning
approaches, such as RNNoise \cite{valin18}, and its extensions like PercepNet \cite{valin20},
Personalized PercepNet \cite{valin21}, etc. Our work focuses on improving
PercepNet due to its excellent ability of improving speech perceptual quality and noise suppression.

PerceptNet \cite{valin20} aims to
enhance fullband (48 kHz sampled) noisy speech with
low-complexity, and has been shown to deliver high-quality speech
enhancement in real-time even when operating on less than 5\% of a CPU core (1.8 GHz Intel i7-8565U CPU). Instead of performing on the Fourier transform bins in state-of-the-art end-to-end SE methods,
PerceptNet features the speech short-time Fourier transform (STFT) spectrum
from 0 to 20 kHz with only 34 bands, according to the human hearing
equivalent rectangular bandwidth (ERB) scale \cite{2012An}, which
greatly lowers the system computational complexity. Together with
the design of pitch filter and envelope postfiltering, PercepNet
can produce high quality enhanced speech.

However, we find that, compared with the enhanced noisy speech with a low SNR,
PercepNet results in much heavier over attenuation (OA) when the input
noisy speech with a relative high SNR. It significantly impairs the perceptual quality of enhanced speech in high SNR condition (even worse than the original noisy speech). This heavier quality impairing
may be due to the inaccurate
estimation of the frequency band gains, and the further speech
enhancing by envelope postfiltering to remove residual noise, since the high SNR noisy speech is actually a clean speech
from human perception. Furthermore, during the PercepNet pipeline, only the speech spectral envelope is enhanced,
the phase of noisy speech is directly used to reconstruct the
target clean speech. All of these mentioned issues might limit the
PercepNet performance.

\begin{figure*}[ht]
  \centering

  \setlength{\belowcaptionskip}{-0.5cm}
  \includegraphics[width=\linewidth]{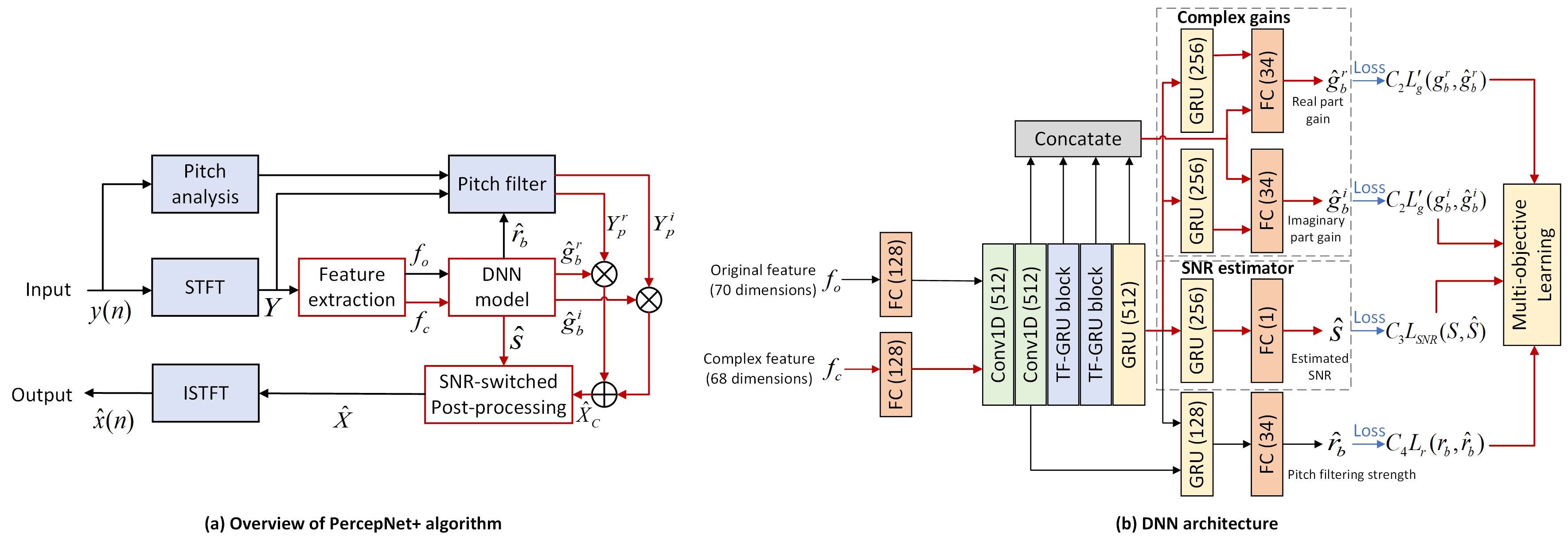}
  \caption{The framework of proposed PercepNet+ algorithm. All dark red blocks and lines
  are our improvements over the original PercepNet\cite{valin20}.
  (a) Overview of PercepNet+; (b) DNN model architecture of PercepNet+.}
  \label{fig:percp}
\end{figure*}

To develop a real-time SE system with better and more robust performance,in this study, we focus on improving the PercepNet to further enhance
its speech denoising ability and achieve better speech perceptual quality.
Four main contributions are: 1) A phase-aware structure is introduced
to leverage the phase information,
by adding the complex subband features as additional
deep network input, and replacing the original
energy gains with subband real and imaginary part gains
for target clean speech construction; 2) To handle the
over attenuation problem and alleviate perceptual
quality impairing of enhanced high SNR noisy speech,
we design an SNR estimator and an SNR-switched post-processing
to control the degree of residual noise removal; 3)
We replace the first two GRU \cite{gru14} layers in PercepNet with
TF-GRU structure to well learn both the time-scale temporal and frequency
dependencies; 4) Based on the above revisions, we finally
propose to learn the complex gains, SNR, the original
pitch filtering strength, together with an OA loss
in a multi-objective training manner to further improve the
SE performance. Compared with PercepNet, our
proposed PercepNet+ achieves absolute 0.19 PESQ \cite{ITU2005ITU} and 2.25\%
STOI \cite{STOI2011} gains on the public VCTK  \cite{vctk16} test set, and
0.15 PESQ and 2.93\% STOI gains on our simulated test set \cite{demo22}.

\vspace{-0.2cm}
\section{PercepNet}
\label{sec:percp}

PercepNet is a perceptually-motivated approach for low-complexity,
real-time enhancement of full-band speech \cite{valin20}.
It extracts various hand-crafted ERB-based subband acoustic features
from 34 triangular spectral bands as the model input.
The model outputs energy gain, $\hat{g_b}$ is then multiplied with the pitch filtered
spectrum of noisy speech to remove the background noise,
where the pitch filter is a comb filter that designed to remove the
noise between pitch harmonics \cite{filter1995}.
The effects of pitch filter in each ERB band is controlled by a pitch filter
strength $\hat{r_b}$. Both $\hat{g_b}$ and $\hat{r_b}$ are automatically
learnt by a deep neural network (DNN), which is
mainly composed of two convolution layers and five GRU layers.
The DNN model utilizes the features of current and three extra future frames to compute its outputs, which makes the PercepNet achieve 30ms look-ahead.
With the help of envelope post-filtering, the denoised speech is
further enhanced. More details can be found in \cite{valin20}.

\vspace{-0.2cm}
\section{Proposed PercepNet+}
\label{sec:ppercep}

As mentioned in Section \ref{sec:intro}, we extend PercepNet
to PercepNet+ in four new aspects: the phase-aware structure, SNR estimator
and SNR-switched post-processing, the multi-objective loss function
and updated TF-GRU blocks. Fig.\ref{fig:percp} illustrates the whole
framework of PercepNet+, and all the dark red blocks and lines
are our improvements over original PercepNet  \cite{valin20}.

\vspace{-0.2cm}
\subsection{Phase-aware Structure}
\vspace{-0.1cm}
\label{subsec:pas}

Because the DNN input features of PercepNet are tied to 34 ERB
bands, as shown in Fig.\ref{fig:percp}, the original
70-dimensional acoustic features $f_o$, are composed of 68-dimensional band-related
features (34 for spectrum energy, 34 for pitch coherence), a pitch period \cite{pperiod1995} and
a pitch correlation \cite{pcorr2013}. For each band $b$, the DNN model
outputs two elements: the energy gain $\hat{g_b}$ and pitch
filter strength $\hat{r_b}$. These features only focus on enhancing the
noisy spectral envelop and pitch harmonics, the
importance of phase information \cite{pim2011} that can affect the
human perception significantly is ignored.

To exploit the phase information in PercepNet+,
we concatenate the real and imaginary part of complex STFT for noisy speech $y(n)$
in each ERB band directly, to form a total 68-dimensional complex feature $f_c$. Then,
as in Fig.\ref{fig:percp}(b), the linear transformed (FC layer) $f_o$ and $f_c$ are finally
concatenated to train the improved DNN model.
Besides adding the complex features, we also replace the original
energy gain with ``complex gains" to pay more attention on phase
as in Fig.\ref{fig:percp}(b). Specifically, we
propose to guide the network to learn real and imaginary part gains, $g_b^r$ and
$ g_b^i$, for re-constructing both target clean speech magnitude and phase spectrum,
and define them as:
\begin{equation}
g_b^r(t) = \frac{{\|X_b^r(t)\|}_{2}}{{\|Y_b^r(t)\|}_{2}}, \ \ g_b^i(t) = \frac{{\|X_b^i(t)\|}_{2}}{{\|Y_b^i(t)\|}_{2}}
  \label{eq1}
\end{equation}
where $X_b (t)$ and $Y_b (t)$ are the complex-valued spectrum of
the clean signal $x(n)$ and its noisy signal $y(n)$ for ERB band $b$
in frame $t$, and $\|\cdot\|_{2}$ means the $L_2$-norm operation.

\vspace{-0.2cm}
\subsection{SNR Estimator and SNR-switched Post-processing}
\label{subsec:snr}

Speech distortion is easy to occur during the process of removing noise,
and it may seriously impair the speech perceptual quality \cite{ZhengPZSL21}.
In PercepNet, this distortion may be due to the inaccurate estimation of
energy gain and the inappropriate design of envelope post-filtering. In PercepNet+, we
propose an SNR estimator and design an SNR-switched post-processing
to alleviate the speech distortion in PercepNet.

\textbf{SNR estimator}: this estimator is inspired by works in
\cite{NicolsonP20,lv21_interspeech}. As shown in Fig.\ref{fig:percp}(b),
it is composed of one GRU and one fully-connected (FC)
layer with sigmoid activation function, and predicts frame-level SNR
under a multi-objective learning
framework to maintain good speech quality. The normalized
ground-truth SNR $S(t)\in [0,1]$ for frame
$t$ of $y(n)$ is defined as:
\begin{equation}
\begin{split}
  S(t) &= \frac{{Q(t) - \mu }}{\sigma } \ \ \ with \\
  Q(t) &= 20{\log _{10}}(X_m(t)/N_m(t))
\end{split}
 \label{eq2}
\end{equation}
where $\mu$ and $\sigma$ are the mean and standard deviation of SNR
$Q(t)$ for the whole noisy speech, $X_m(t)$ and $N_m(t)$
represent the magnitude spectrum of clean speech and noise respectively.

\textbf{SNR-switched MMSE-LSA post-processing}: Although post-processing module is proved very effective in
removing residual noise \cite{9414062,post2}, it may impair the perceptual quality
of test samples with almost no noise as we found in our experiments.
Therefore, in our PercepNet+, as shown in Fig.\ref{fig:percp}(a),
the predicted SNR $\hat{S}$ of each frame
is used to control whether the post-processing module should be performed.
We call this strategy as SNR-switched post-processing. If $\hat{S}$
is greater than a predefined threshold,
the spectrum ${\hat X_c}$ enhanced by $\hat g_b^r$ and $\hat g_b^i$ will directly be
the final output. Otherwise, ${\hat X_c}$ will be further enhanced
by post-processing to remove residual noise.

In addition, we find that, the conventional MMSE-LSA
\cite{lsa85} based post-processing has achieved remarkable results in recent
end-to-end SE systems \cite{lv21_interspeech,li21g_interspeech}. Therefore,
in PercepNet+, we also replace the original envelop postfiltering with
MMSE-LSA in our SNR-switched post-processing module as follows:
\begin{equation}
  G(t) = \textmd{MMSE-LSA}(\xi (t),\gamma (t))
  \label{eq5}
\end{equation}
\begin{equation}
  \hat X(t) = G(t)*\hat X_c(t)
  \label{eq6}
\end{equation}
where $G(t)$ is the MMSE-LSA frame-level gain,  $\hat X_c(t)$ is the spectrum enhanced
by complex gains of frame $t$,  $\hat X(t)$ is the final enhanced clean speech as
in Fig.\ref{fig:percp}(a), and $\xi (t)$, $\gamma (t)$ are priori and posterior frame-level SNR that
defined as in \cite{lsa85}.

\vspace*{-0.1cm}
\subsection{Multi-objective Loss Function}
\label{subsec:multi}

The original loss function $L_{P}$ of DNN model in PercepNet has two parts:
the loss of energy gain $L_g$ and pitch filter strength $L_r$ as defined:
\begin{equation}
\begin{split}
  {L_g} &= \sum\limits_b {(g_b^\lambda }  - \hat g_b^\lambda {)^2} + {C_1}\sum\limits_b {(g_b^\lambda }  - \hat g_b^\lambda {)^4}\\
  {L_r} &= \sum\limits_b {((1 - {r_b}} {)^\lambda } + {(1 - {\hat r_b})^\lambda }{)^2} \\
  {L_{P}} &= \alpha {L_g} + \beta {L_r}
\end{split}
  \label{eq11}
\end{equation}
where $g_b, \hat{g_b}, r_b, \hat{r_b}$ are the ground-truth and DNN predicted ERB-based band energy gain
and pitch filter strength, respectively. $C_1,\lambda,\alpha,\beta$ are tuning parameters.

Besides the SNR-switched post-processing, results in \cite{eskimez2021personalized}
showed an asymmetric loss $L_{OA}$ that proposed in \cite{wang20z_interspeech} is effective
in alleviating the over attenuation issue. Therefore, we adapt it to $L_g$ to address
the quality impairing in high SNR condition as follows:
\begin{equation}
h(x) = \left\{ \begin{array}{l}
0,{\quad}if{\;}x \le 0,\\
x,{\quad}if{\;}x > 0,
\end{array} \right.
  \label{eq12}
\end{equation}
\begin{equation}
  {L_{OA}}(g_b,\hat g_b) = {\left| {h(g_b - \hat g_b)} \right|^2}
  \label{eq13}
\end{equation}
\begin{equation}
  {L'_g} = \delta {L_g} + (1 - \delta ){L_{OA}}.
  \label{eq14}
\end{equation}

In PercepNet+, instead of using $L_P$,  we use Eq.(\ref{eq14}) to
measure the difference between estimated $\hat g_b^r$, $\hat g_b^i$ and
their ground-truth respectively. Together considering the original $L_r$, and
mean-square-error (MSE) loss of SNR ${L_{SNR}}({S},\hat S )$, the
DNN model of PercepNet+ is finally joint trained using the following
overall multi-objective loss function $L_{P+}$:
\begin{equation}
\begin{split}
  L_{P+} = {C_2}{L'_g}({g_b^r},{\hat g_b^r}) + {C_2}{L'_g}({g_b^i},{\hat g_b^i}) \\
    + {C_3}{L_{SNR}}({S} ,\hat S ) + {C_4}{L_r}(r_b,\hat r_b)
  \label{eq15}.
\end{split}
\end{equation}
where $C_2,C_3, C_4$ are tuning loss weight parameters.

\vspace*{-0.1cm}
\subsection{TF-GRU Block}
\label{subsec:tfgru}

PercepNet models the temporal dependency at time-scale with GRU layers. Inspired by
\cite{7404793}, we employ another GRU layer to model the frequency-wise evolution of spectral
patterns. Specifically, as shown in Fig.\ref{fig:percp}(b),
we replace two GRU layers in PercepNet with two proposed TF-GRU blocks,
each TF-GRU is composed of one time-GRU (TGRU) layer and one frequency-GRU
(FGRU) layer. The FGRU is used to browse the frequency band features so that the frequency-wise
dependency is summarized, and the TGRU is used to summarize time-wise dependency.
The outputs of both TGRU and FGRU are then concatenated to form the final
TF-GRU output.The parameter quantity of one TF-GRU is adjust to be consistent with one GRU layer in original PercepNet.

\vspace{-0.2cm}
\section{Experimental Setup}
\label{sec:es}

\subsection{Datasets}
\label{subsec:data}

The training data used in original PercepNet \cite{valin20}
is not publicly available, so we use the public dataset that used to train
RNNoise model in \cite{valin18} as our training set. The clean speech
data is from McGill TSP speech database \cite{data1} and
the NTT Multi-Lingual Speech Database for Telephonometry \cite{data2}.
Various sources of noise are used, including computer fans, office, crowd, airplane, car,
train, construction. Totally, we have 6 hours clean speech and 4 hours noise data, which is
far less than the 120 hours speech plus 80 hours noise data used in
PercepNet\cite{valin20}.
The training pairs are simulated by dynamically mixing noise and speech
with a random SNR ranges from -5 to 20dB.
Half of the speech are convoluted with room impulse response (RIR) coming from RIR\underline{~}NOISES set
\cite{data3}.

Two evaluation sets are used to examine the proposed techniques, one is
the public noisy VCTK test set \cite{vctk16} with
824 samples from 8 speakers. The other
is a test set named D-NOISE \cite{demo22}, which is simulated by ourselves, with
SNR ranging from -5 to 20dB. D-NOISE consists of
108 samples with the speech data from WSJ0 \cite{wsj0} dataset
and noise data comes from RNNoise demo website \cite{rdemo18}, including
the office, kitchen, cars, street and babble noises.

\vspace*{-0.2cm}
\subsection{Configurations}
\label{subsec:cfg}

All training and test data are 48kHz sampled. The frames are extracted using a
Vorbis window \cite{vorbis04} of size 20 ms and an overlap of 10 ms. Batch size is set to 32.
Adam optimizer \cite{KingmaB14} is used, and the initial earning rate is set to 0.001.
We set the loss function weights, $\lambda,\alpha,\beta,\delta$ to 0.5, 4.0, 1.0 and 0.7,
respectively, and $C_{1},C_{2},C_{3},C_{4}$  to 10, 4, 1 and 1 respectively.
The DNN network layer parameters of PercepNet+  are presented in Fig.\ref{fig:percp}(b).
To make results comparable, all other configurations are the same as
the original PercepNet in \cite{valin20} and RNNoise in \cite{valin18}. Both PESQ and STOI are used to measure the speech quality and intelligibility repecetively.

\vspace{-0.2cm}
\section{Results and discussion}

\subsection{Baseline}

\vspace{-10pt}
\begin{table}[h]

  \caption{PESQ and STOI(\%) on VCTK test set. The training data of model 2, 4 are 190 hours
  less than that of model 1,3.}
  \label{tab:base}
  \centering
  \begin{tabular}{clcc}
    \toprule
\textbf{ID} &   \textbf{Model}      &\quad\;\textbf{PESQ}&\quad\;\textbf{STOI}  \\
    \midrule
- &    Noisy                                      & \quad\;1.97   & \quad\ 92.12                    \\
1  &  RNNoise \cite{valin18}                                & \quad\;2.29       & \quad\ -                \\
2  &  RNNoise                          & \quad\;2.23       & \quad\ 92.74                \\
3  &  PercepNet \cite{valin20}                                  & \quad\;2.54   & \quad\ -            \\
4  &  PercepNet                               & \quad\;2.46& \quad\ 93.43                    \\
    \bottomrule
  \end{tabular}
  \vspace{-10pt}
\end{table}

Both the RNNoise (open-sourced) and its extension-PercepNet (not open-sourced)
are taken as our baselines.
Table \ref{tab:base} presents the comparison results on
VCTK test set. Model 1 and 3 are the published results in PercepNet\cite{valin20},
in which the models were trained on non-public 120 hours speech plus 80 hours noise data,
while model 2 and 4 are our implemented RNNoise and PercepNet models that
trained on only the 6 hours speech and 4 hours noise data.
It's clear that, PercepNet outperforms significantly its original
RNNoise, and the PESQ scores of our models are only slightly
worse than the ones in \cite{valin20}, even there is an extremely big training
data size gap (190 hours) between our models and model 1,3.
Therefore, we think that our implementation of PercepNet
is correct and can be taken as the baseline of our PercepNet+.

\begin{figure}[t]

  \setlength{\belowcaptionskip}{-0.3cm}
  \centering
  \includegraphics[width=\linewidth]{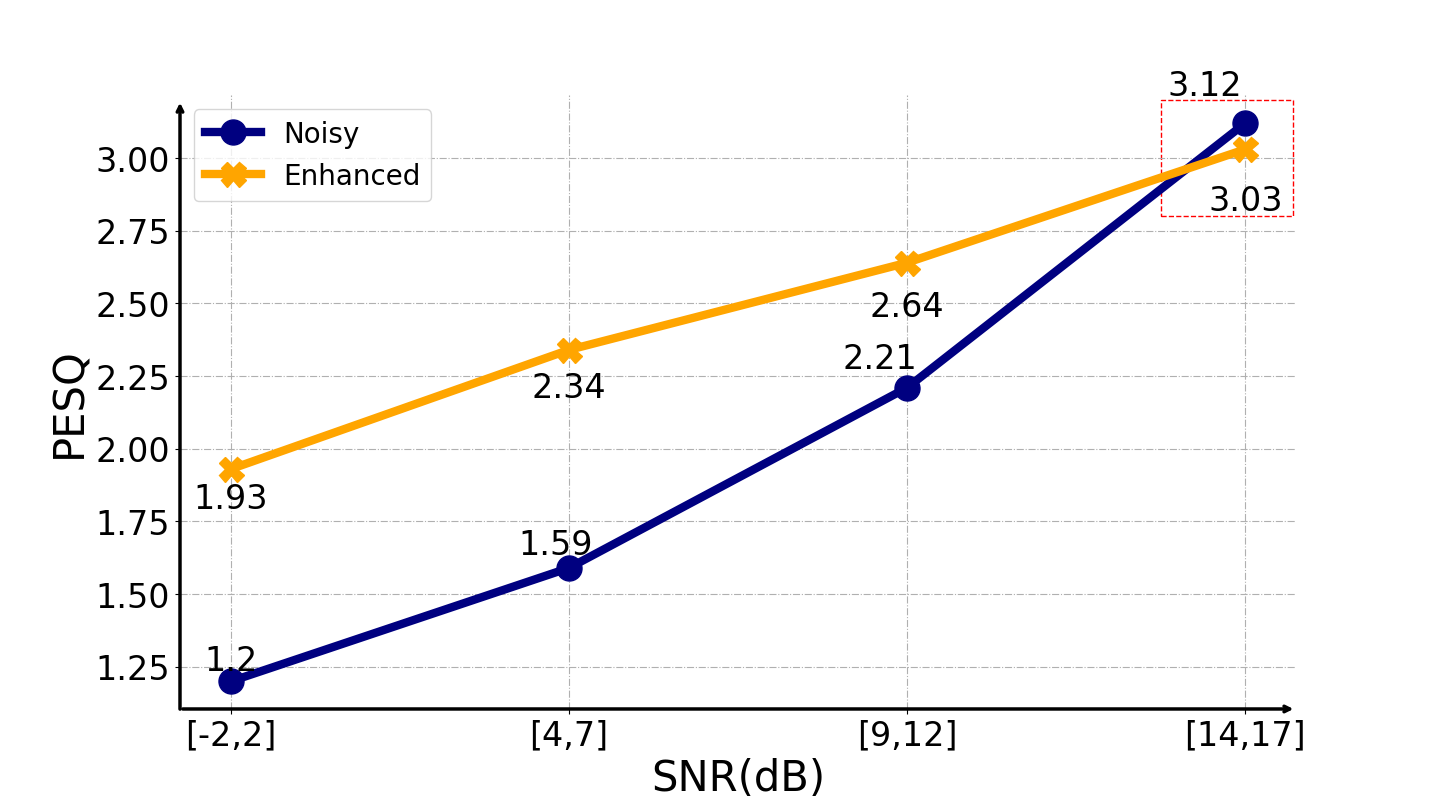}
  \caption{PESQ of noisy and PecepNet enhanced VCTK test samples with different SNR ranges.}
  \label{fig:oa}
\end{figure}

\begin{figure}[t]
  \setlength{\abovedisplayskip}{-0.1cm}
  \setlength{\belowdisplayskip}{-0.5cm}
  \centering
  \includegraphics[width=\linewidth]{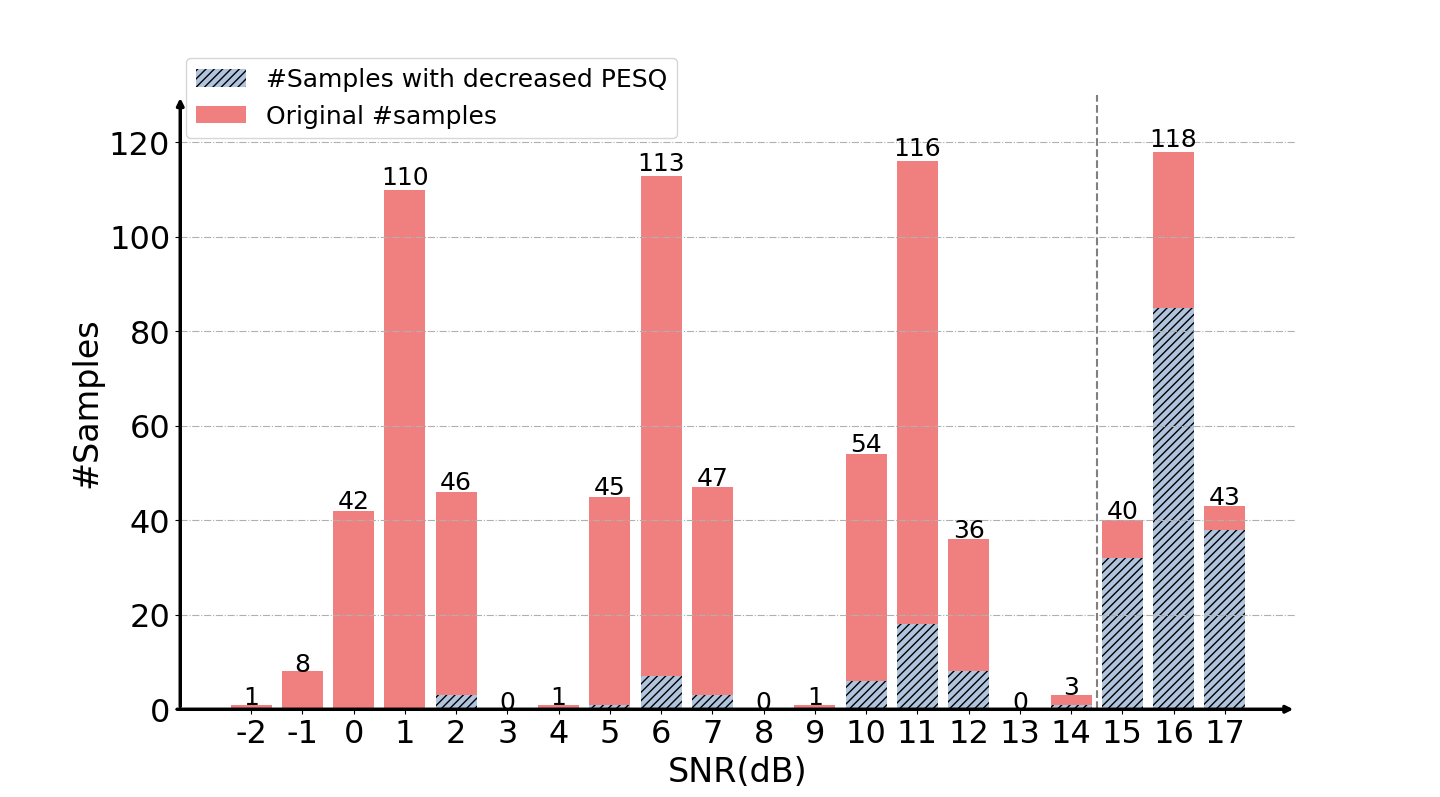}
  \caption{Detail samples distribution with decreased PESQ after enhancement
  by PercepNet in different SNR conditions of VCTK test set.}
  \label{fig:sample}
  \vspace{-15pt}
\end{figure}

Moreover, a detail baseline results analysis are presented in
Fig.\ref{fig:oa} and \ref{fig:sample}, in which the whole VCTK test set
is divided into 4-level SNR ranges to observe the PecepNet denoising performance
and behaviors in different SNR conditions. From Fig.\ref{fig:oa},
we find that the PESQ of samples with SNR$>$14dB is decreased after
enhancement. Meanwhile, in Fig.\ref{fig:sample},
comparing the histogram of red and its corresponding light blue parts,
it's clear that most PESQ decrease occurs in higher SNR conditions,
specifically, there are total 202 samples with decreased PESQ and
76.35\% of them are with SNR greater than 14dB. It indicates that
the original PercepNet has heavier OA or fails to perform well under high SNR conditions.
Therefore, in this study, we take the 14dB as our proposed SNR-switched post-processing
threshold.

\vspace{-0.2cm}
\subsection{Results of PercepNet+}
\label{subsec:rper}

Table \ref{tab:ppvd} shows the performance comparison of all the
proposed techniques in PercepNet+ on both VCTK and our simulated D-NOISE test
sets. Compared with PercepNet, the proposed PercepNet+ significantly
improves the PESQ from 2.46 to 2.65, and STOI from 93.43\% to 95.68\%
on VCTK test set. Specifically, the additional complex features and gains
lead to an absolute increase of 0.08 and 1.11\% in terms of PESQ and
STOI. With the help of SNR estimator, we obtain 0.04 PESQ and
0.27\% STOI improvements. When the SNR-switched post-processing (PP),
and over attenuation loss are applied, the PESQ and STOI reach to 2.62 and 95.49\%.
Finally, we see the updated TF-GRU results further performance gains.
Moreover, on D-NOISE test set, we achieve consistent performance gains
from all the proposed techniques of PercepNet+, an overall 0.15 PESQ and
2.93\% STOI gains are obtained. In addition, the proposed PercepNet+
has 8.5M trainable parameters, an increase of 0.5M compared with PercepNet, 
and Real Time Factor (RTF) is equal to 0.351, which is tested on a machine with an Intel(R) Xeon(R) CPU E5-2650 v2@2.60GHz in single thread.
Therefore, we can draw the conclusion that, PercepNet+ has greatly
surpassed PercepNet without significantly increasing the parameters
of neural network.

\vspace{-5pt}
\begin{table}[h]
  \caption{Various models’ PESQ and STOI(\%) on both VCTK and D-NOISE test set.}
  \label{tab:ppvd}
  \centering
  \scalebox{0.95}{
  \begin{tabular}{lcc}
    \toprule
    \multirow{2}{0.1\textwidth}{\textbf{Model}}&\textbf{VCTK}&\textbf{D-NOISE}\\
    \cmidrule(lr){2-2}\cmidrule(lr){3-3}
    ~&\textbf{PESQ \;\; STOI}&\textbf{PESQ \;\;STOI}\\
    \midrule
    Noisy                  &\ 1.97 \quad\ 92.12 & \ 2.10  \quad\ 86.53                     \\
    RNNoise                &\ 2.23 \quad\ 92.74 & \ 2.33  \quad\ 88.27                    \\
    PercepNet              &\ 2.46 \quad\ 93.43 & \ 2.57  \quad\ 90.53                   \\
    \;+\,Complex Features  &\ 2.51 \quad\ 93.88 & \ 2.60  \quad\ 91.28                        \\
    \;\;+\,Complex Gains   &\ 2.54 \quad\ 94.54 & \ 2.63  \quad\ 92.19                 \\
    \;\;\;+\,SNR Estimator &\ 2.58 \quad\ 94.87 & \ 2.65  \quad\ 92.41              \\
    \;\;\;\;+\,SNR-switched PP &\ 2.61 \quad\ 95.40 & \ 2.69 \quad\ 93.11         \\
    \;\;\;\;\;+\,OA Loss                 &\ 2.62 \quad\ 95.49 & \ 2.70 \quad\ 93.12                         \\
    \;\;\;\;\;\;+\,TF-GRU (PercepNet+)                &\ \textbf{2.65} \quad\ \textbf{95.68} & \ \textbf{2.72} \quad\ \textbf{93.46}\\
    \bottomrule
  \end{tabular}}
  \vspace{-10pt}
\end{table}

\vspace{-5pt}
\subsection{Performance of SNR-sensitive techniques}
\label{subsec:est}

We further investigate the effectiveness of proposed
OA loss and SNR-switched PP in solving the speech perceptual quality
impairing issue after enhancement in high SNR condition.
Results of both two VCTK sub test sets (“$>$14dB” and “$\leq$14dB”)
are shown in Table \ref{tab:snrs}. Comparing the PESQs in first
two lines, we see the PercepNet does impair the
perceptual quality with high SNR, which is consistent as observed
in Fig.\ref{fig:oa}. However, in PercepNet+, we see this issue is
effectively alleviated by either the proposed OA loss or the
SNR-switched PP. When both two techniques are applied,
the performance is further slightly improved, without
impairing the speech perceptual quality in low SNR condition.
\begin{table}[h]
  \caption{PESQ scores of PercepNet+ with (or without) over attenuation(OA) loss and SNR switch on VCTK sub test sets}
  \label{tab:snrs}
  \centering
  \begin{tabular}{llll}
    \toprule
    \textbf{Model} &\textbf{$\leq $14dB}&\textbf{\textgreater14dB}&\textbf{Overall}              \\
    \midrule
    Noisy               & \;\;\;1.59                         & \;\;\;3.12     & \;\;\;1.97                  \\
    PercepNet     & \;\;\;2.28                         & \;\;\;3.03     & \;\;\;2.46                  \\
    PercepNet+           & \;\;\;\textbf{2.49}                        & \;\;\;\textbf{3.14}     & \;\;\;\textbf{2.65}\\
    $-$SNR Switch         & \;\;\;\textbf{2.49}           & \;\;\;3.11     & \;\;\;2.64               \\
    $-$OA Loss     & \;\;\;2.48                  & \;\;\;3.13     & \;\;\;2.64            \\
    \;\;\;\;$-$SNR Switch    & \;\;\;2.48     & \;\;\;3.03     & \;\;\;2.62                      \\
    \bottomrule
  \end{tabular}
  \vspace{-10pt}
\end{table}

\vspace{-5pt}
\section{Conclusions}
\label{sec:con}

In this paper, we propose PercepNet+, by extending the
high quality, real-time and low-complexity PercepNet in several
aspects to further improve speech enhancement, including the phase-aware
structure to learn the phase information, two SNR-sensitive improvements
to maintain speech perceptual quality while removing noise, the
updated TF-GRU to model time and frequency scale dependencies
simultaneously, and the multi-objective learning for further
system performance improvements. Importantly, the heavy
speech perceptual quality impairing by the original PercepNet
in high SNR conditions has been well solved due to the proposed
OA loss and SNR-switched post-processing. Experimental
results show that the proposed PercepNet+ has significantly
outperformed the original PercepNet in both PESQ and STOI.
Some noisy and enhanced samples, including the D-NOISE test set can be found from \cite{demo22}.

\bibliographystyle{IEEEtran}
\bibliography{isbib}
\end{document}